\documentclass[aps, onecolumn, superscriptaddress]{revtex4-2}

\usepackage[utf8]{inputenc} 
\usepackage[T1]{fontenc}    
\usepackage{hyperref}       
\usepackage{url}            
\usepackage{booktabs}       
\usepackage{amsfonts}       
\usepackage{nicefrac}       
\usepackage{microtype}      
\usepackage{lipsum}
\usepackage{amsmath}
\usepackage{amsfonts}
\usepackage{amssymb}
\usepackage{indentfirst}
\usepackage{graphicx}
\usepackage{subfigure}
\usepackage{tikz}
\usepackage{pgfplots}
\usepackage{changepage}
\usepackage{xcolor}

\newcommand{\bra}[1]{\ensuremath{\left\langle#1\right|}}
\newcommand{\ket}[1]{\ensuremath{\left|#1\right\rangle}}
\newcommand{\bracket}[2]{\ensuremath{\left\langle#1 \vphantom{#2}\right| \left. #2 \vphantom{#1}\right\rangle}}

\begin{document}

\title{Phonon mediated non-equilibrium correlations and entanglement between distant semiconducting qubits}

\author{Di Yu}
\affiliation{Institute of Fundamental and Frontier Sciences, University of Electronic Science and Technology of China, Chengdu 610054, China}
\affiliation{Yingcai Honors College, University of Electronic Science and Technology of China, Chengdu 611731, China}
\author{Zhi-Meng Guo}
\affiliation{Yingcai Honors College, University of Electronic Science and Technology of China, Chengdu 611731, China}
\author{Guang-Wei Deng}
\email{gwdeng@uestc.edu.cn}
\affiliation{Institute of Fundamental and Frontier Sciences, University of Electronic Science and Technology of China, Chengdu 610054, China}
\affiliation{CAS Key Laboratory of Quantum Information, University of Science and Technology of China, Hefei 230026, China}

\date{\today}

\begin{abstract}
We theoretically study the non-equilibrium correlations and entanglement between distant semiconductor qubits in a one-dimensional coupled-mechanical-resonator chain. Each qubit is defined by a double quantum dot (DQD) and embedded in a mechanical resonator. The two qubits can be coupled, correlated and entangled through phonon transfer along the resonator chain. We calculate the non-equilibrium correlations and steady-state entanglement at different phonon-phonon coupling rates, and find a maximal steady entanglement induced by a population inversion. The results suggest that highly tunable correlations and entanglement can be generated by phonon-qubit hybrid system, which will contribute to the development of mesoscopic physics and solid-state quantum computation.
\end{abstract}

\maketitle

\section{Introduction}
Semiconducting quantum dot is one of the candidates for quantum computing \cite{Loss1998}. One challenge in this field is how to achieve long-range coupling and entanglement between qubits. During the past decade, pioneering experiments have been implemented to reach this goal \cite{DelbecqPRL,frey2012dipole,Petersson2012,DengPRL2015,DengNL2015,Mi2017science,Stockklauser2017PRX,Mi2018nature,Samkharadze2018science,Borjans2020nature}. The general idea is to couple quantum dots to a Bosonic resonator such as a microwave (photon) resonator \cite{ChildressPRA}, which have been widely studied in superconducting qubit systems \cite{Wallraff:Nature,Xiang2013Hybrid}. However, this new kind of hybrid-circuit quantum electrodynamics \cite{Burkard2020nrp}, which require superconductor–semiconductor heterogeneous integration, have posed new challenges to nanotechnologies. For example, after heterogeneous integration, quality factors of the superconducting microwave resonator will be significantly reduced \cite{DelbecqPRL,frey2012dipole,Petersson2012,DengPRL2015,DengNL2015}. Therefore, it is necessary to handle samples very carefully or design more complex structures to achieve strong coupling regimes \cite{Mi2017science,Stockklauser2017PRX,Mi2018nature,Samkharadze2018science,Borjans2020nature}. Actually, in solid-state systems, the gate-defined quantum dots can be strongly coupled to mechanical vibrations (or phonons) \cite{Lassagne2009,Steele2009}. Therefore, it is possible to engineer nano-electro-mechanical system (NEMS) as another potential candidate for quantum-dot-based hybrid quantum electrodynamics. Previously, coherent quantum phonon dynamics \cite{Yiwen2017science}, phonon Fock states control \cite{Yiwen2018nature} and remote superconducting qubit entanglement \cite{Bienfait2019science} have been realized by acoustic-wave resonator. 

In this paper, we consider carbon nanotube (CNT) mechanical resonator as a model system to study the coupling dynamics between phonon modes and quantum dots. Current technology has enabled the CNT resonator with a quality factor of $5\times10^6$ \cite{Moser2014Nanotube} and an eigenfrequency of GHz level \cite{Chaste2011High,Laird2012,Xinhe2018nr}. These properties suggest that a CNT resonator is possible to serve as a quantum phonon bus at low temperature. Double quantum dot (DQD) can be realized by local gates beneath or above the CNT  \cite{van2002Electron,Biercuk2005,Laird2015rmp}. Various experiments have studied the electron-phonon interactions between quantum dots and the vibration modes in CNT resonators \cite{Lassagne2009,Steele2009,Eichler2012,Meerwaldt2012,Benyamini2014Real,Deng2016Strongly,Shu-Xiao2016nanoscale,Dong2017nl,Xinhe2018nr,Xinhe2020nr} and a coupling strength of 320 MHz between double quantum dot (DQD) and mechanical state has been reported \cite{Khivrich2019Nanomechanical}. Based on these results and the fact that phonons can be coherently transferred in a coupled mechanical resonator chain \cite{Hajime2013,Faust2013,Deng2016Strongly,Luo2018Strong,zhang2020coherent}, here we aim to theoretically study the generation of steady and temporal entanglement states between distant DQD qubits, mediated by phonons in coupled CNT resonators. Starting from a simplified Hamiltonian and calculating the non-equilibrium correlations between two DQDs, we report a maximal steady entanglement because of population inversion between the second and third eigenstates, and we also propose a method to calculate the evolution of phonon-mediated temporal entanglement.

\section{Model}
\subsection{General Hamiltonian and Liouvillian}
\begin{figure}[htbp]
	\centering
	\includegraphics[width=15cm]{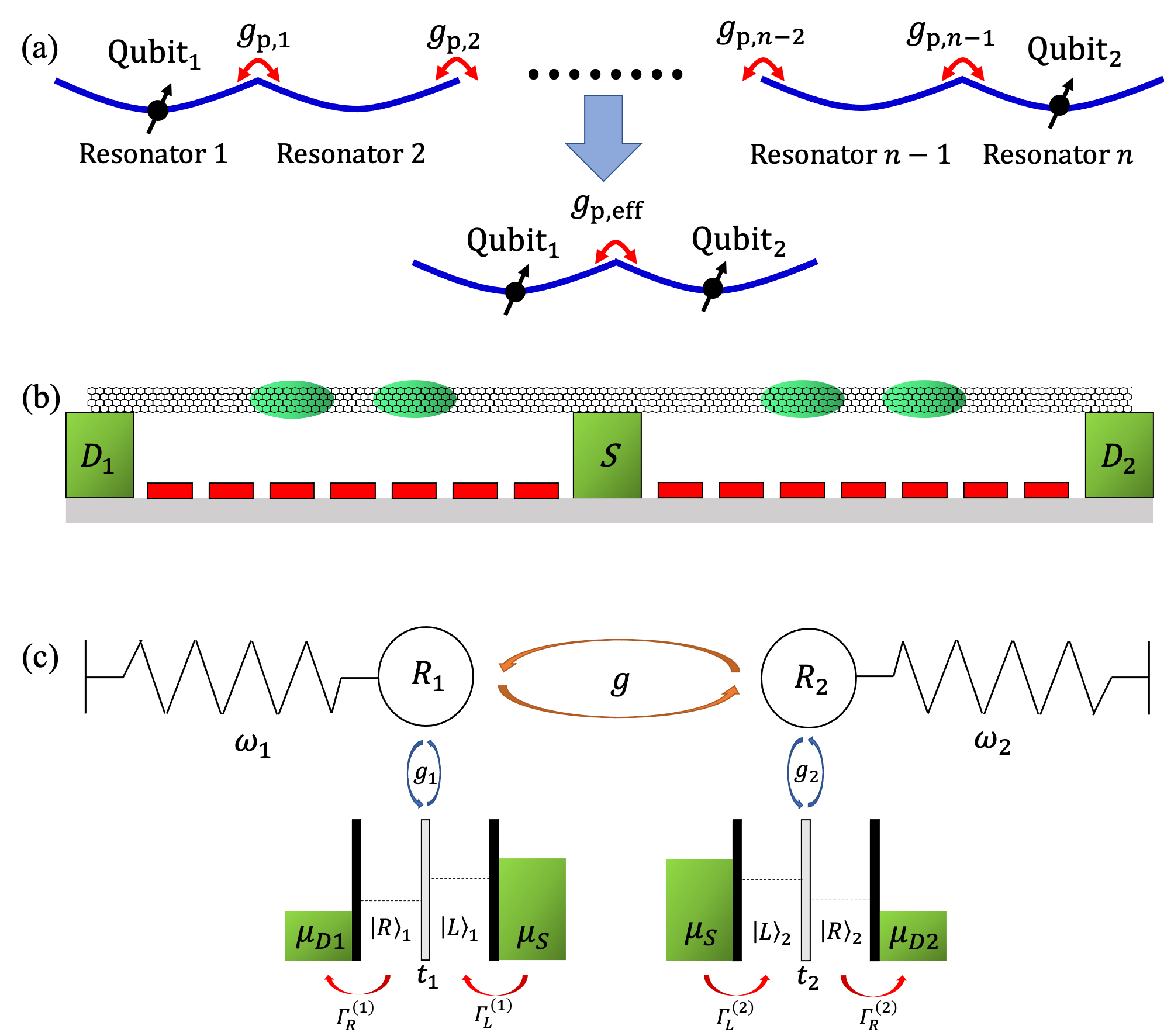}
	\caption{Model schematic of the coupling system. (a) A resonator chain containing $n$ resonators and two qubits are coupled to the first resonator and $n$th resonator, respectively. Coupling strength between $i$th resonator and $(i+1)$th resonator is noted as $g_{p,i}$. The coupling between the first and $n$th resonators can be simplified to an effective $g_{p,eff}$. (b) A carbon nanotube suspended on three electrodes, forming two mechanical resonators. Electro-gates (red) under the two resonators are used to tune the mechanical resonance frequency and define the two double quantum dots (green), respectively. (c) Equivalent model of the coupled system, which consists of two DQDs and two mechanical resonators. The eigenfrequency of the two resonators $R_1$ and $R_2$ are $\omega_1$ and $\omega_2$, respectively, and they are coupled with a coupling strength of $g$. Each resonator couples to an embedded DQD with a strength of coupling $g_i$. Here we use $\ket{L}$ to denote the state of a DQD with an extra electron trapped in the quantum dot near the source electron, where the chemical potential is $\mu_S$. While the state of a DQD $\ket{R}$ indicates the extra electron locates in the quantum dot adjacent to the drain electrode, where the chemical potential is $\mu_{Di}$. The inner coupling strength of the $i$th DQD is $t_i$, and both DQDs are coupled to electron reservoirs with tunnel rates $\Gamma_R^{(i)}$ and $\Gamma_L^{(i)}$. \label{figure.1}}
\end{figure}

As shown in Fig. \ref{figure.1}, the model system is composed of two DQDs embedded in two CNT resonators, respectively, and the two resonators are coupled as well. Each DQD is coupled to two electronic reservoirs on either side of the DQD. We assume that the capacitance of each DQD is suitable that no more than one electron is possible to tunnel in and out of the DQD, and the charge qubit is defined by the superposition of the two electron states $\ket{L}_i,\ket{R}_i (i=1,2)$ \cite{van2002Electron}. To take electrons tunnel between DQDs and reservoirs into consideration, we define another quantum state $\ket{0}$ as zero electron in both DQD. Now we can write the Hamiltonian of this system composed of two DQDs and two CNT resonators, which we called 'location representation' for convenience.

The Hamiltonian consists of four parts, the Hamiltonian of DQDs, resonators, coupling between DQDs and resonators, and coupling between the two resonators, denoted by $H_{el}, H_{res}, H_{el-res}, H_{res-res}$, respectively. Denoted the detuning and strength of tunneling coupling of each DQD by $\epsilon_i$, $t_i$ respectively, we then write the total Hamiltonian of the two DQD as 
\begin{equation}
\label{E1}
	H_{el}=\sum_{i=1}^{2} \left(\frac{\epsilon_i}{2}\sigma_z^{(i)}+t_i \sigma_x^{(i)}\right),
\end{equation}
where $\sigma_x^{(i)},\sigma_z^{(i)}$ are Pauli matrix of the \textit{i}th DQD. Assuming the mode of resonator related to coupling has an eigenfrequency of $\omega_i$ for the \textit{i}th CNT, the total vibration Hamiltonian of both resonators reads $H_{res}=\sum_{i=1}^{2} \omega_i a_i^\dagger a_i$. By quantizing the classical Hamiltonian of coupling between resonator and DQD in \cite{Khivrich2019Nanomechanical}, we get
\begin{equation}
\label{E2}
	H_{el-res}=\sum_{i=1}^{2} \hbar g_i \sigma_z^{(i)}\left(a_i^\dagger+a_i\right),
\end{equation}
where $g_i$ is the strength of coupling between \textit{i}th DQD and \textit{i}th resonator. Actually, it is similar with the coupling Hamiltonian between a DQD and a microwave resonator \cite{Contreras2013Non}. Denoting the coupling strength between the two nanotubes by $g$, and using rotating wave approximation (RWA), we write the Hamiltonian of coupling between resonators as
\begin{equation}
\label{E3}
	H_{res-res}=\hbar g\left(a_1^\dagger a_2 + a_1 a_2^\dagger\right).
\end{equation}
And the total Hamiltonian of the system reads
\begin{equation}
\label{E4}
	H_{sys}=H_{el}+H_{res}+H_{el-res}+H_{res-res}.
\end{equation}

Now we introduce another representation, which seems more simple in mathematics. Considering the two eigenstates of each DQD, denoted by $\ket{\uparrow}_i,\ket{\downarrow}_i$ for the state with higher energy $\frac{\Omega_i}{2}$ and the other with lower energy $-\frac{\Omega_i}{2}$, and introducing auxiliary parameters $A_i=\frac{g_i \epsilon_i}{\Omega_i},B_i=-\frac{2g_i t_i}{\Omega_i}$, it is easy to find $g_i \sigma_z^{(i)}=A_i \tilde{\sigma}_z^{(i)}+B_i\tilde{\sigma}_x^{(i)}$ and $H_{el}=\sum_{i=1}^{2}\frac{\Omega_i}{2}\tilde{\sigma}_z^{(i)}$, where $\sigma_x^{(i)}$ and $\sigma_z^{(i)}$ is both Pauli matrix in $\ket{\uparrow}_i,\ket{\downarrow}_i$ representation. In the representation defined by $\ket{\uparrow}_i,\ket{\downarrow}_i$ and phonon number states, $H_{el-res}=\sum_{i=1}^{2} \hbar(A_i \tilde{\sigma}_z^{(i)} + B_i \tilde{\sigma}_x^{(i)})(a_i^\dagger + a_i)$. Considering that the $H_{res}$ and $H_{res-res}$ remain unchanged in this 'energy representation', we can obtain the total Hamiltonian in another representation.

Having obtained the total Hamiltonian of the system, we then move to Liouvillian of the system. Assuming that each electronic reservoir in the left side of a DQD is with a suitably high fermi level, while the other one in right side is with a suitably low one, so that only electron transfer from the left reservoir to the left dot and from the right dot to the right reservoir dominates (see Fig. 1(c)), each with tunnel rate $\Gamma_L^{(i)},\Gamma_R^{(i)}$, respectively. Denoting the dissipation rate of each resonator by $\kappa_i$, the Liouvillian of the system at a temperature of 0 K reads
\begin{equation}
\begin{aligned}
\label{E5}
	L(\rho)=&-\frac{i}{\hbar}[H_{sys},\rho]\\
	&+\sum_{i=1}^{2} -\frac{\Gamma_L^{(i)}}{2}\left[s_{L,i}s_{L,i}^\dagger \rho-2s_{L,i}^\dagger\rho s_{L,i}+\rho s_{L,i}s_{L,i}^\dagger\right]\\
	&-\frac{\Gamma_R^{(i)}}{2}\left[s_{R,i}^\dagger s_{R,i}\rho-2s_{R,i} \rho s_{R,i}^\dagger+\rho s_{R,i}^\dagger s_{R,i}\right]\\
	&-\frac{\kappa_i}{2}\left[a_i^\dagger a_i \rho -2a_i \rho a_i^\dagger+\rho a_i^\dagger a_i\right],
\end{aligned}
\end{equation}
where $s_{L,i}$ is the annihilate operator of the left quantum dot of the \textit{i}th DQD. Here the master equation describes evolution of the system coupled to environment with $\frac{d}{dt}\rho (t)=L(\rho (t))$.

\subsection{Effective Hamiltonian}
To extract the effective indirect coupling from the total Hamiltonian of system, we calculate the effective Hamiltonian of the two coupled qubits \cite{Soliverez1981General,Ren2019A}. For simplicity, only the states with no more than one phonon in total is taken in consideration, and the calculation is performed to the third order. The effective Hamiltonian in the location representation reads
\begin{equation}
\label{E6}
	H_{eff}=\sum_{i=1}^{2}\left(\frac{2}{\epsilon_i}\sigma_z^{(i)}+t_{i,eff}\sigma_x^{(i)}\right)+J_z'\sigma_z^{(1)}\sigma_z^{(2)}-\sum_{i\neq j}J_{xz,ij}' \sigma_z^{(i)} \sigma_x^{(j)},
\end{equation}
where
\begin{equation}
\begin{aligned}
\label{E7}
	&t_{i,eff}=t_i\left[1+\frac{2g_i^2}{(\Omega_i+\omega_i)(\Omega_i-\omega_i)}\right],\\
	&J_z'=\sum_{i\neq j}\frac{2g_1 g_2 g t_i^2}{\Omega_i^2}\left[\frac{1}{(\Omega_i-\omega_i)(\Omega_i-\omega_j)}+\frac{1}{(\Omega_i+\omega_i)(\Omega_i+\omega_j)}\right]+\frac{g_1g_2g\epsilon_1\epsilon_2}{\omega_1\omega_2\Omega_1\Omega_2}\sum_{i\neq j}\frac{\Omega_i\epsilon_j}{\Omega_j \epsilon_i},\\
	&J_{xz,ij}'=\frac{g_1 g_2 g \epsilon_jt_j}{\Omega_j^2}\left[\frac{1}{(\Omega_j-\omega_j)(\Omega_j-\omega_i)}-\frac{1}{(\Omega_j+\omega_j)(\Omega_j+\omega_i)}\right]+\frac{2g_1g_2g\epsilon_1\epsilon_2}{\omega_1\omega_2\Omega_1\Omega_2}\frac{\Omega_it_j}{\Omega_j\epsilon_i}.
\end{aligned}
\end{equation}
The interaction Hamiltonian is composed by the Ising interaction Hamiltonian and the XZ exchange interaction Hamiltonian, when $t_i \ll \left|\epsilon_i\right|$, the XZ exchange interaction Hamiltonian would dominate.

In the energy representation, the effective Hamiltonian reads
\begin{equation}
\label{E8}
	H_{eff}=\sum_{k=1}^{2}\frac{\tilde{\Omega}_{k,eff}}{2}\tilde{\sigma}_z^{(k)}+\xi_k\tilde{\sigma}_x^{(k)}+\alpha \tilde{\sigma}_z^{(1)}\tilde{\sigma}_z^{(2)}+\sum_{k\neq l=1,2}\beta_{kl}\tilde{\sigma}_z^{(k)}\tilde{\sigma}_x^{(l)}+\gamma \tilde{\sigma}_x^{(1)} \tilde{\sigma}_x^{(2)},
\end{equation}
where
\begin{equation}
\begin{aligned}
\label{E9}	
	&\tilde{\Omega}_{k,eff}=\Omega_k \left(1+\frac{2B_k^2}{(\Omega_k-\omega_k)(\Omega_k+\omega_k)}\right),\\
	&\xi_k=-\frac{A_k B_k \Omega_k}{(\Omega_k-\omega_k)(\Omega_k+\omega_k)},\\
	&\alpha=\frac{2A_1 A_2 g}{\omega_1 \omega_2},\\
	&\beta_{kl}=A_k B_l g\left[\frac{1}{\omega_k\omega_l}+\frac{1}{2(\Omega_l+\omega_l)(\Omega_l+\omega_k)}+\frac{1}{2(\Omega_l-\omega_l)(\Omega_l-\omega_k)}\right],\\
	&\gamma=\frac{B_1 B_2 g}{2}\sum_{i\neq j=1,2}\left[\frac{1}{(\Omega_i+\omega_i)(\Omega_i+\omega_j)}+\frac{1}{(\Omega_i-\omega_i)(\Omega_i-\Omega_j)}\right].
\end{aligned}
\end{equation}
Similar with the effective Hamiltonian in the location representation, when $t_i \ll \left|\epsilon_i\right|$, the XZ exchange Hamiltonian would dominates.

The effective Hamiltonian, derived from the full Hamiltonian, describes only the indirect coupling between qubits. It is the key for us to explain the generation mechanism of steady entanglement. Though the effective Hamiltonian is simple, it is available only when the number of phonons in a resonator is far lower than 1, and it also requires that the direct or indirect coupling between the two qubits or between the qubits and the resonators is weak $(\Omega_i\neq \omega_j)$, which excludes circumstances of interest such as strong coupling condition and resonant condition. To study these excluded conditions analytically, we apply RWA approximation on the general Hamiltonian. 

\subsection{Physical parameters}
Noting that the parameters $\epsilon_i,\ t_i,\ \Gamma_{L/R}^{(i)}$ are tunable by means of gate voltages, we focus on the eigenfrequency of the resonators $\omega_i$, dissipation rate $\kappa_i$, the strength of coupling between DQDs and resonators $g_i$, and coupling between resonators $g$. Coupling strength between DQD and resonator depends on odd symmetry of vibration modes, here we consider the second mode of nanotubes, which accompanying an eigenfrequency ranging from 100 MHz to several GHz \cite{Deng2016Strongly,Chaste2011High}, and a typical value for Q factor of the second mode of the nanotube is $10^4$ \cite{Khivrich2019Nanomechanical,Deng2016Strongly}. The strength of coupling between DQD and resonator can reach 320 MHz \cite{Khivrich2019Nanomechanical}. Considering state-of-the-art strongly coupled phonon cavities, the coupling between resonators can realize a coupling strength between from 0.01 to 0.1 times the resonator eigenfrequency \cite{Luo2018Strong,zhang2020coherent}. Here we use typical parameters as $\omega_1=1,\ g_1=g_2=0.03\sim0.06,\ g=0.01\sim0.1,\ \kappa_1=\kappa_2=10^{-4}$ in our simulation, where $\omega_1$ plays the role of unit. For convenience, we define $\hbar=1$ below. 

\section{Steady generation of entanglement}
\subsection{Weak coupling between resonators ($g=0.01$)}
\begin{figure}[htbp]
\includegraphics[width=18cm]{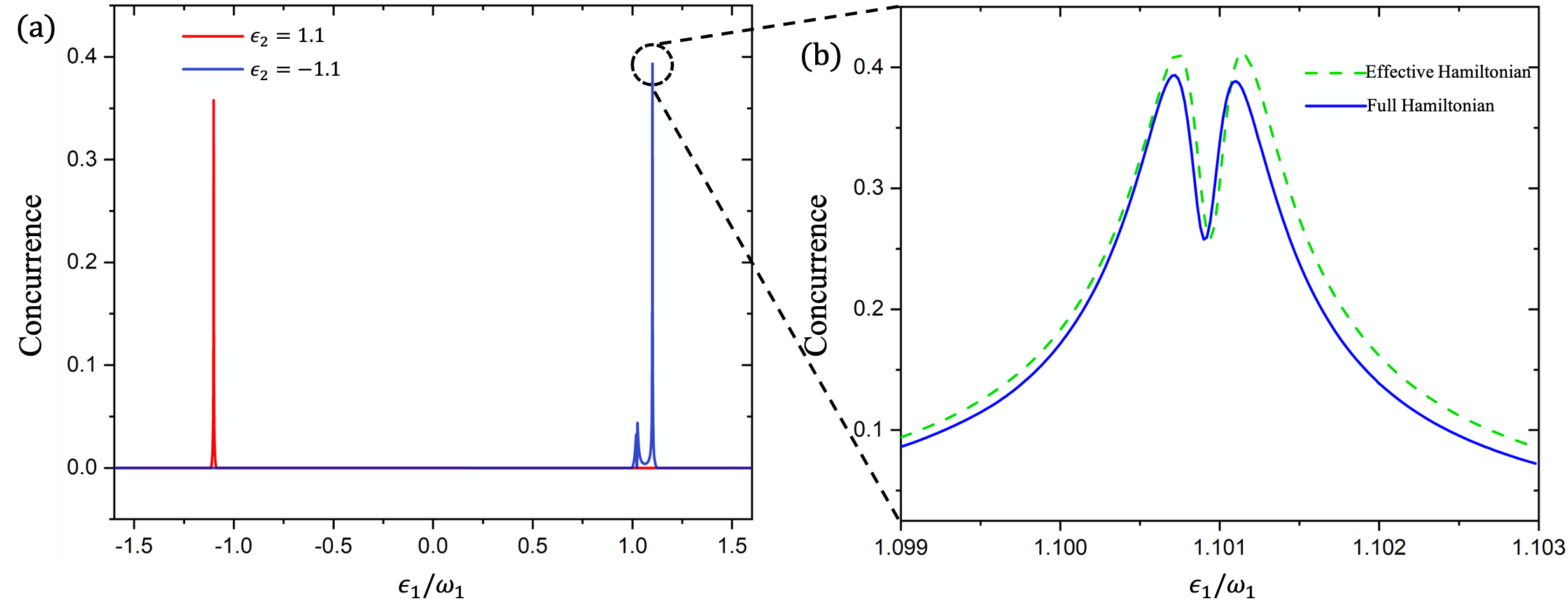}
\caption{(a) Concurrence of the two qubits based on the full Hamiltonian, where $\epsilon_1$ is scanned and $\epsilon_2$ is set to be $\pm 1.1$ respectively, while $\omega_1=1,\omega_2=1.1,t=0.2,g_1=g_2=0.03,g=0.01,\kappa=10^{-4}$ and $\Gamma=10^{-4}$. Two main peaks and a sub peak are observed around $\epsilon_1+\epsilon_2=0$ and $\Omega_1=\omega_2$ respectively. The sub peak origins from the indirect resonance between the first DQD and the second resonator. (b) Zoom near the maximal concurrence in panel (a), and a comparison of the maximal concurrence calculated with the effective Hamiltonian and that calculated with the full Hamiltonian.\label{figure.2}}
\end{figure}

We first study the steady entanglement in a situation with weak coupling ($g=0.01<g_1, g_2$) between resonators. For simplicity, we assume that $t_1=t_2=t,g_1=g_2,\kappa_1=\kappa_2=\kappa$ and $\Gamma_{L,i}=\Gamma_{R,i}=\Gamma$. Here we employ the concurrence to represent the degree of entanglement of the two qubits \cite{Hill1997Entanglement}. The result of simulation is shown in Fig .\ref{figure.2}.

The two maximal peaks are both located near $\epsilon_1=-\epsilon_2$ because $\left|\epsilon_1\right|=\left|\epsilon_2\right|$ can lead to the degeneracy of energy levels of two DQDs as $\Omega_1=\Omega_2$ and in turn leads to a larger effective coupling between DQDs. In order to explain the opposite sign for the concurrence peaks $\epsilon_1=-\epsilon_2$, we consider a DQD level isolated from the resonator and only coupled to the source and drain electronic reservoir with the assumption that $t\ll \left|\epsilon\right|$. If $\epsilon<0$, the steady state will be approximately equal to $\ket{\downarrow}$; when $\epsilon>0$, the steady-state is approximately equal to $\ket{\uparrow}$, partially because relatively small $t$ would localize the excess electron in the left quantum dot for steady states, noting that $\ket{\downarrow}$ (when $\epsilon<0$) and $\ket{\uparrow}$ (when $\epsilon>0$) are close to $\ket{L}$. A complete discussion on the steady state of an isolated DQD is given in appendix.B. Now we can understand why the steady states approximate to $\ket{\uparrow\downarrow}$ or $\ket{\downarrow\uparrow}$ when $\epsilon_1=-\epsilon_2$. In addition, as we have discussed in the section of effective Hamiltonian, the XZ exchange interaction dominates the qubit-qubit interaction in energy representation when $t_i \ll \left|\epsilon_i\right|$. As a result, states $\ket{\uparrow\downarrow}$ and $\ket{\downarrow\uparrow}$ should produce the maximum entanglement when $\epsilon_1=-\epsilon_2$. In contrast, state $\ket{\uparrow\uparrow}$ or $\ket{\downarrow\downarrow}$ is approximately the steady state when $\epsilon_1=\epsilon_2$, producing almost no entanglement between the two qubits. This is why the maximum entanglement comes with the condition $\epsilon_1=-\epsilon_2$. The phenomenon observed here is similar with that from the coupling between two DQDs and a microwave resonator \cite{Contreras2013Non}.

\begin{figure}
\includegraphics[width=18cm]{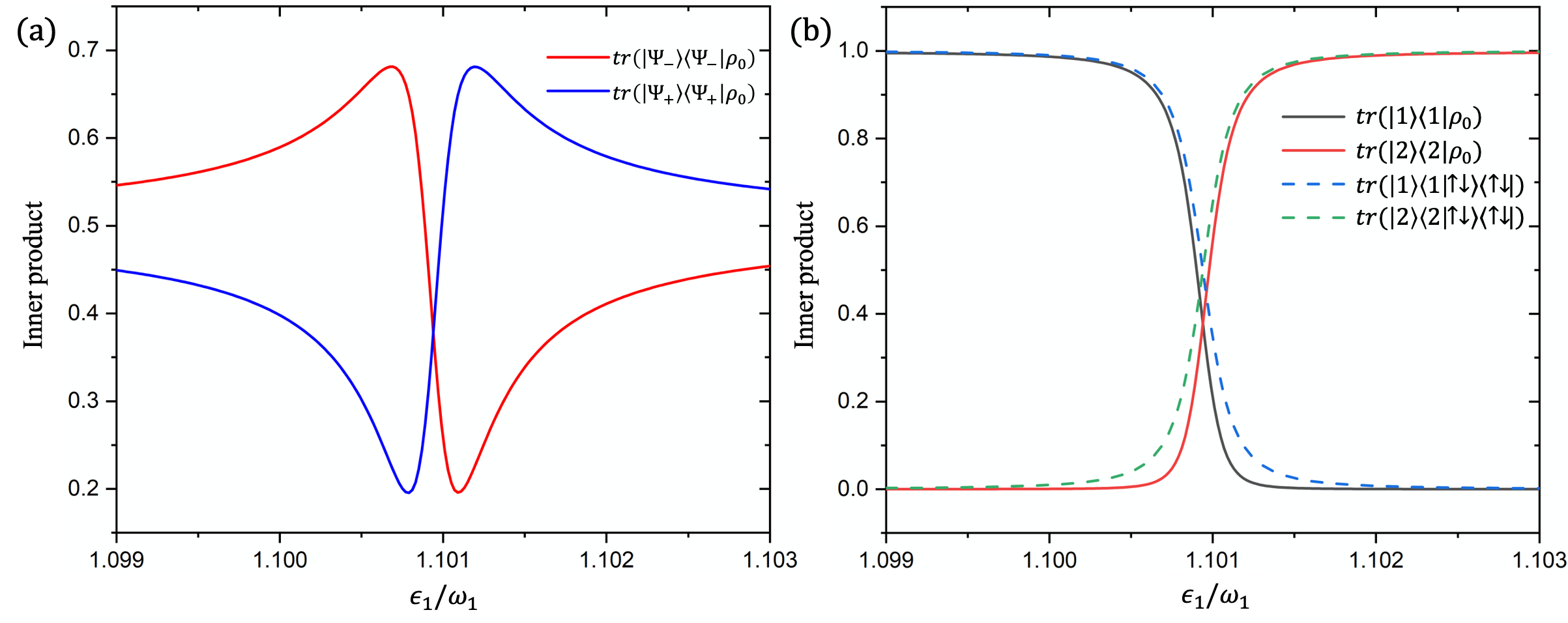}
\caption{$tr(\ket{\Psi_-}\bra{\Psi_-}\rho_0)$ and $tr(\ket{\Psi_+}\bra{\Psi_+}\rho_0)$ are shown in (a), based on the effective Hamiltonian. The simulation uses the same parameters as in Figure.\ref{figure.2}. As we expected, the double peaks of concurrence correspond to $\ket{\Psi_-}$ and $\ket{\Psi_+}$ respectively, and the transfer of the steady state between the two Bell states leads to the valley of concurrence. (b) Inner product versus the offset $\epsilon_1$. $tr(\ket{1}\bra{1}\rho_0)$ and $tr(\ket{2}\bra{2}\rho_0)$, as well as $tr(\ket{1}\left\langle1|\uparrow\downarrow\right\rangle\bra{\uparrow\downarrow})$ and $tr(\ket{2}\left\langle2|\uparrow\downarrow\right\rangle\bra{\uparrow\downarrow})$ are calculated, based on the effective Hamiltonian. The simulation parameters the same as Fig .\ref{figure.2}.\label{figure.3}}
\end{figure}

Considering that the effective Hamiltonian will be used in the simulation of the maximal concurrence, a test of the accuracy on the effective Hamiltonian is needed here. A comparison of the maximal concurrence calculated with the effective Hamiltonian and that calculated with the full Hamiltonian is shown in Fig. \ref{figure.2}(b), and the effective Hamiltonian shows enough accuracy. Fig. \ref{figure.2}(b) renders a double-peak structure in the diagram of concurrence versus $\epsilon_1$ near the maximal concurrence peak, which is closely connected to the generation of the maximal steady concurrence. Intuitively, the maximum entanglement should occur at resonance of the two qubits, which indicates maximal concurrence. A natural explanation of the counter-intuitive double-peak structure is that the steady state of the subsystem composed of two qubits changes approximately from a Bell state to another Bell state when $\epsilon_1$ changes from producing a concurrence peak to generating the other concurrence peak, and the steady state superposed by two Bell states leads to the valley of concurrence between the two concurrence peaks. This draws a question, which states superpose to produce the concurrence peaks and the concurrence valley? To answer this question, we need to think about the degeneration of the state $\ket{\uparrow\downarrow}$ and the state $\ket{\downarrow\uparrow}$ when $\epsilon_1=1.1$. The steady state of the subsystem consistes of two qubits, which should be a superposition state when $\epsilon_1=1.1$. Here $\ket{\uparrow\downarrow}$ is degenerate with the state $\ket{\downarrow\uparrow}$ when $\epsilon_1=1.1$. Because we have limited to basis composed of Bell states, the two Bell states $\ket{\Psi_+}=(\ket{\downarrow\uparrow}+\ket{\uparrow\downarrow})/\sqrt{2}$ and $\ket{\Psi_-}=(\ket{\downarrow\uparrow}-\ket{\uparrow\downarrow})/\sqrt{2}$, superposed by both $\ket{\uparrow\downarrow}$ and $\ket{\downarrow\uparrow}$, can superpose to produce the steady state when $\epsilon_1$ is in proximity to $1.1$. Simulation result of $tr(\ket{\Psi_+}\bra{\Psi_+}\rho_0)$ and $tr(\ket{\Psi_-}\bra{\Psi_-}\rho_0)$ shown in Fig .\ref{figure.3}(a) proves our inference, where $\rho_0$ is the density operator of the steady state. To conclude, steady superposed by $\Psi_-$ and $\Psi_+$ leads to the concurrence valley between two concurrence peaks.

Nevertheless, there is a relatively large difference between the steady state and the corresponding Bell states $\ket{Psi_-}$ and $\ket{\Psi_+}$ even at the concurrence peaks in Fig .\ref{figure.3}(a), limiting the maximum concurrence to about 0.4, as shown in Fig .\ref{figure.2}. For further understanding of this difference, we consider inner product of the second eigenstate $\ket{1}\bra{1}$, the third eigenstate $\ket{2}\bra{2}$. The steady state are named by $tr(\ket{1}\bra{1}\rho_0)$ and $tr(\ket{2}\bra{2}\rho_0)$, as shown in Fig .\ref{figure.3}(b). The result shows the steady state is approximately a superposition of $\ket{1}$ and $\ket{2}$, and increasing $\epsilon_1$ would convert $\rho_0$ from $\ket{1}$ to $\ket{2}$, which raises two questions: 1. How does the converting relate to the transfer of the steady state between Bell states? 2. What causes such converting? 

Considering the weak coupling condition, $\ket{1}\approx\ket{\uparrow\downarrow}$ and $\ket{2}\approx\ket{\downarrow\uparrow}$ when $\epsilon_1<\epsilon_2$ and vice versa. $\ket{\uparrow\downarrow}$ and $\ket{\downarrow\uparrow}$ become degenerate at $\epsilon_1=-\epsilon_2=1.1$, thus we speculate that degeneracy of $\ket{1}$ and $\ket{2}$ takes place at the valley of concurrence namely at $\epsilon_1\approx 1.101$. The anti-crossing of energy level makes $\ket{1}$ close to $\ket{\Psi_-}$ and makes $\ket{2}$ close to $\ket{\Psi_+}$ here, then the transfer between $\ket{1}$ and $\ket{2}$ leads to the valley of concurrence.

To illuminate the reason of the population inversion, we consider the relation between $\ket{1},\ket{2}$ and $\ket{\uparrow\downarrow},\ket{\downarrow\uparrow}$, where the calculated $tr(\ket{1}\bracket{1}{\uparrow\downarrow}\bra{\uparrow\downarrow})$ and $tr(\ket{2}\bracket{2}{\uparrow\downarrow}\bra{\uparrow\downarrow})$ are shown in Fig .\ref{figure.3}(b). The steady state of two DQDs is $\ket{\uparrow\downarrow}$ if isolated from each other, then the anti-crossing leads to swift transfer of $\ket{1}$ from $\ket{\uparrow\downarrow}$ to $\ket{\downarrow\uparrow}$, and that of $\ket{2}$ from $\ket{\downarrow\uparrow}$ to $\ket{\uparrow\downarrow}$. In this case, by scanning $\epsilon_1$ near the resonance between qubits, a population inversion involved with $\ket{1}$ and $\ket{2}$ happens.

\subsection{Strong coupling between resonators ($g=0.1$)}
\begin{figure}
\includegraphics[width=10cm]{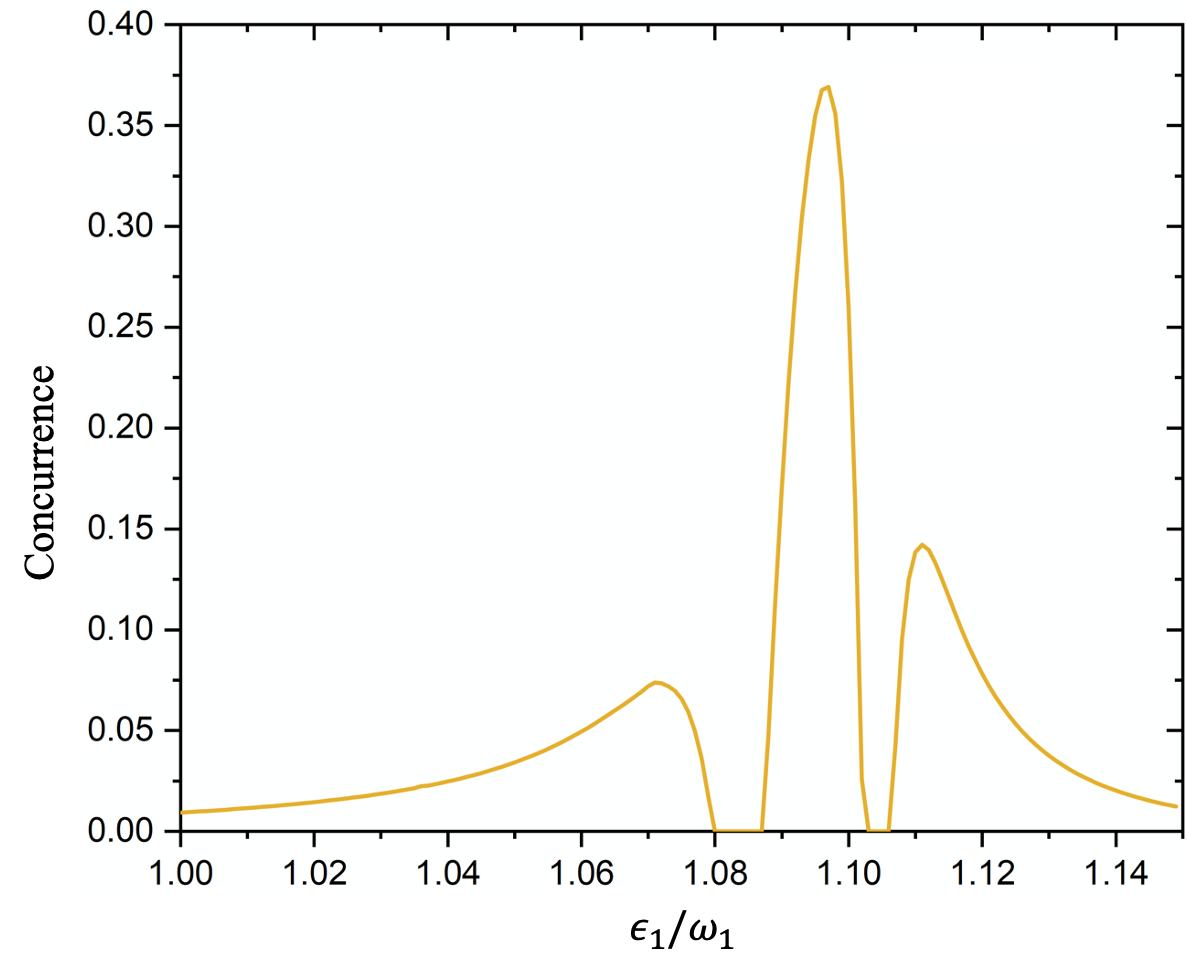}
\caption{Concurrence of qubits versus $\epsilon_1$ based on the full Hamiltonian, here $g=0.1$, other parameters remain the same as that in Fig .\ref{figure.3}. Concurrence with $\epsilon_1$ in other range is almost zero, and only the part with $\epsilon_1 \in (1.00,1.15)$ is plotted here. \label{figure.4}}
\end{figure}

Now we move to the case where coupling between resonators is strong ($g=0.1>g_1, g_2$), here the indirect coupling between qubits greatly increases because of the strong resonator-resonator coupling, which leads to larger third-order indirect coupling, so the effective Hamiltonian is no longer valid, and the discussion above is non-duplicated here. Keeping other parameters unchanged, the relationship between $\epsilon_1$ and concurrence is plotted in Fig .\ref{figure.4}. We find three peaks structure of the concurrence by analyzing the intermediate parameter setup using the effective Hamiltonian (See more details in appendix.D). We conclude that the two concurrence peaks on the right side split from the maximal concurrence peak with $g$ increasing, and another peak on the left side evolves from the sub-peak in Fig .\ref{figure.2}(a). 

We analyze the sub peak, which ranges from $\epsilon_1\approx$ 1.02 to 1.08 (the center of the valley). Because of the strong coupling between resonators, the eigenstates of resonators are mixed, thus the sub peak originally suitable for $\Omega_1=\omega_2$ is now owing to be suitable for $\Omega_1=\omega'$, where $\hbar\omega'$ is the eigenenergy of coupled resonators. For coupled resonators with Hamiltonian $H=\omega_1 a_1^\dagger a_1 + \omega_2 a_2^\dagger a_2 + g(a_1^\dagger a_2 + a_1 a_2^\dagger)$, the second and third eigenenergy split from $\omega_1$ and $\omega_2$ to 
\begin{equation}
\label{E10}
	\frac{1}{2}\left(\omega_1+\omega_2 \pm \sqrt{\Delta^2+4g^2}\right),\Delta=\omega_1-\omega_2,
\end{equation}
here $E_3\approx \Omega_1$ leads to $\epsilon_1\approx 1.09$, resulting in the movement of the sub peak.

Another phenomenon is the broadening of each concurrence peak observed in Fig .\ref{figure.4} in comparison with Fig .\ref{figure.2}. A qualitative interpretation can be made with the Heisenberg uncertainty. Denoting the full width at half maximum for the maximal concurrence peak for weak resonator-resonator coupling condition with $\Delta E_1$, and that for strong resonator-resonator coupling condition by $\Delta E_2$, and assuming that a period of the exchange between two qubits $\Delta t\propto g^{-1}$, we have
\begin{equation}
\label{E11}
	\Delta E_1 g_1^{-1}\sim\Delta E_2 g_2^{-1}.
\end{equation}
As a result, it should meet that $\Delta E_1\sim 0.1\Delta E_2$, which is consistent with the simulation when $\Delta E_1\sim 0.001$ and $\Delta E_2\sim 0.01$. Further discussion of the strong resonator-resonator coupling mechanism is difficult here and could be studied in future work.

\section{Dynamical generation of entanglement}
It is difficult to apply analytical method to solve the eigenstates and eigenenergy of the total Hamiltonian as the non-conserversion coupling term $H_{el-res}=\sum_{i=1}^{2} \hbar g_i \sigma_z^{(1)} (a_i^\dagger + a_i)=\sum_{i=1}^{2} \hbar\left(A_i \tilde{\sigma}_z^{(i)} + B_i \tilde{\sigma}_x^{(i)}\right)(a_i^\dagger+a_i)$, which leads to a non-conserversion of the calculation amount of the simulation. We assume $\left|\Omega_i-\omega_i\right| \ll \left|\Omega_i + \omega_i\right|$, then RWA on the coupling term is valid, this gives
\begin{equation}
\label{E12}
	H_{el-res}=\sum_{i=1}^{2}\hbar\left(A_i \tilde{\sigma}_z^{(i)} + B_i \tilde{\sigma}_x^{(i)}\right)(a_i^\dagger + a_i)\approx \sum_{i=1}^{2} \hbar B_i\left(\tilde{\sigma}_-^{(i)}a_i^\dagger+\tilde{\sigma}_+^{(i)}a_i\right).
\end{equation}
With this approximation, the total Hamiltonian is commutable with the operator $N=\sum_{i=1}^{2} a_i^\dagger a_i + \frac{1}{2}\left(\tilde{\sigma}_z^{(i)}+1\right)$. In this case, the total Hamiltonian is reduced to multiple isolated subspaces, we then discuss the dynamical generation of entanglement using this approximation.

For simplicity, we assume that $\omega_1=\omega_2=\Omega_1=\Omega_2$, and $\kappa=\Gamma=0$. We reduced the Hilbert space describing the full system into span$(\ket{\uparrow\downarrow 00},\ket{\downarrow\uparrow 00},\ket{\downarrow\downarrow 10},\ket{\downarrow \downarrow 01})$. The reduced Hamiltonian reads
\begin{equation}
\label{E13}
	H=
	\begin{pmatrix}
	0&0&B&0\\
	0&0&0&B\\
	B&0&0&g\\
	0&B&g&0
	\end{pmatrix}
	.
\end{equation}
Setting the initial state to be $\ket{\uparrow\downarrow 00}$, and solving the Schrodinger equation, we get the dynamical evolution of the state as
\begin{equation}
\label{E14}
\frac{\sqrt{E_1 E_2}}{E_1 - E_2}
\begin{pmatrix}
-\sqrt{\frac{E_2}{E_1}}\cos{(E_1 t)}+\sqrt{\frac{E_1}{E_2}}cos{(E_2 t)}\\
-i\sqrt{\frac{E_2}{E_1}}sin{(E_1 t)}+i\sqrt{\frac{E_1}{E_2}}sin{(E_2 t)}\\
-i sin{(E_1 t)}+i sin{(E_2 t)}\\
-cos{(E_1 t)}+cos{(E_2 t)}
\end{pmatrix}
,
\end{equation}
where $E_{1,2}=\frac{1}{2}(-g\pm \sqrt{g^2+4B^2})$ is the first and the second eigenenergy of the reduced Hamiltonian. How to find the maximally entangled state for a superposition between $\ket{\uparrow\downarrow 00}$ and $\ket{\downarrow\uparrow 00}$?

Considering that the number of phonon sinusoidally oscillate with a period $T=\frac{2\pi}{\sqrt{g^2+4B^2}}$. Here only the moment when the number of phonon is zero need to be considered, namely $t_i=\frac{2\pi i}{\sqrt{g^2+4B^2}} (i=0,1,2...)$. Denoting $E_1t_i=\theta_i$, the states in these moments read
\begin{equation}
\label{E15}
\begin{pmatrix}
\cos{\theta_i}\\
i\sin{\theta_i}\\
0\\
0
\end{pmatrix}
(i=1,2,3,...).
\end{equation}
Here the condition for Bell state formation is $\left|cos{\theta_i}\right|=\left|\sin{\theta_i}\right|$. If the maximally entangled state forms at $t=t_1$, the relationship as follow should be meet:
\begin{equation}
\label{E16}
	\left|\frac{g}{B}\right|=\frac{2}{\sqrt{15}}\ or\ \frac{6}{\sqrt{7}}.
\end{equation}
This is the condition for the fastest generation of maximally entangled state. Typical parameters meeting the relation above could be $g=0.05$, $g_1=g_2=0.06$ and $t=0.1837$, which are reachable by tuning $\epsilon_i$ and $t_i$. Here we have showed the dynamical evolution with the resonant condition $\omega_1=\omega_2=\Omega_1=\Omega_2$, and studied the condition for the generation of maximally entangled state.

All calculations in this paper was performed with QuTip \cite{QuTip}.

\section{Conclusion}
In conclude, we have studied the concurrence of two biased qubits indirectly coupled with each other and intermediated by mechanical resonators. We focus on both steady-state and dynamical evolution of entanglement generation. For steady state with weak coupling between resonators, the maximal concurrence owes to the degeneracy of the second and the third eigenstates of the effective Hamiltonian. When the two qubits are in resonance, a concurrence valley of the steady state counter-intuitively occurs, because two Bell states superpose to produce the steady state. For strong coupling between resonators, a movement of the sub concurrence peak and the broadening of concurrence peaks are observed. The former is due to the mix of eigenstates of resonators, while the latter can be explained with the Heisenberg uncertainty relationship. We also studied the dynamical evolution of the entanglement, and find that proper parameter setup could enable quick generation of a non-phonon entangled state. Our results could benefit the field of semiconducting quantum computing.

\appendix
\section{Derivation of the effective Hamiltonian}
Using the third order perturbation method and considering states with no more than one phonon in total in two resonators, we have
\begin{equation*}
	H_{eff}=PH_0P+\nu^{(1)}+\nu^{(2)}+\nu^{(3)},
\end{equation*}
where $P$ is the projector operator to the non-phonon subspace. Denoting $\sum_{x\notin Im_P, y\in Im_P} \frac{1}{E_y-E_x} P_x A P_y$ by $(A)$, the perturbation terms read \cite{Ren2019A}
\begin{equation*}
	\nu^{(1)}=PVP=0,
\end{equation*}
\begin{equation*}
	\begin{aligned}
		\nu^{(2)}
		&=PV(V)P\\
		&=\sum_{k=1}^{2}\left(-\frac{A_k^2}{\omega_k}+\frac{B_k^2 \omega_k}{(\Omega_k-\omega_k)(\Omega_k+\omega_k)}\right)I_4 + \sum_{k=1}^{2}\frac{B_k^2 \Omega_k}{(\Omega_k-\omega_k)(\Omega_k+\omega_k)}\tilde{\sigma}_z^{(k)}-\sum_{k=1}^{2}\frac{A_kB_k\Omega_k}{(\Omega_k-\omega_k)(\Omega_k+\omega_k)}\tilde{\sigma}_x^{(k)},
	\end{aligned}
\end{equation*}
\begin{equation*}
	\begin{aligned}
		\nu^{(3)}
		&=P[V(V(V))-V((V)V)]P\\
		&=\frac{2A_1A_2g}{\omega_1 \omega_2}\tilde{\sigma}_z^1 \tilde{\sigma}_z^2 + \frac{B_1B_2g}{2}\sum_{i\neq j=1,2}\left[\frac{1}{(\Omega_i+\omega_i)(\Omega_i+\omega_j)}+\frac{1}{(\Omega_i-\omega_i)(\Omega_i-\omega_j)}\right]\tilde{\sigma}_x^{(1)}\tilde{\sigma}_x^{(2)}\\
		&\ \ \ + \sum_{k\neq l=1,2}A_kB_lg\left[\frac{1}{\omega_k \omega_l}+\frac{1}{2(\Omega_l+\omega_l)(\Omega_l+\omega_k)}+\frac{1}{2(\Omega_l-\omega_l)(\Omega_l-\omega_k)}\right]\tilde{\sigma}_z^{(k)}\tilde{\sigma}_x^{(l)}.
	\end{aligned}
\end{equation*}
So, the total effective Hamiltonian reads
\begin{equation*}
	\begin{aligned}
		H_{eff}&=\sum_{k=1}^{2} \frac{\Omega_k}{2}\tilde{\sigma}_z^{(k)}+\left(\sum_{k=1}^{2} -\frac{A_k^2}{\omega_k}+\frac{B_k^2 \omega_k}{(\Omega_k-\omega_k)(\Omega_k+\omega_k)}\right)I_4 + \sum_{k=1}^{2}\frac{B_k^2 \Omega_k}{(\Omega_k-\omega_k)(\Omega_k+\omega_k)}\tilde{\sigma}_z^{(k)}\\
		&\ \ \ -\sum_{k=1}^{2}\frac{A_kB_k\Omega_k}{(\Omega_k-\omega_k)(\Omega_k+\omega_k)}\tilde{\sigma}_x^{(k)} + \frac{2A_1A_2g}{\omega_1 \omega_2}\tilde{\sigma}_z^{(1)} \tilde{\sigma}_z^{(2)}\\
		&\ \ \ + \frac{B_1B_2g}{2}\sum_{i\neq j=1,2}\left[\frac{1}{(\Omega_i+\omega_i)(\Omega_i+\omega_j)}+\frac{1}{(\Omega_i-\omega_i)(\Omega_i-\omega_j)}\right]\tilde{\sigma}_x^{(1)}\tilde{\sigma}_x^{(2)}\\ 
		&\ \ \ + \sum_{k\neq l=1,2}A_kB_lg\left[\frac{1}{\omega_k \omega_l}+\frac{1}{2(\Omega_l+\omega_l)(\Omega_l+\omega_k)}+\frac{1}{2(\Omega_l-\omega_l)(\Omega_l-\omega_k)}\right]\tilde{\sigma}_z^{(k)}\tilde{\sigma}_x^{(l)}.
	\end{aligned}
\end{equation*}
Dropping the $I_4$ term irrelevant to dynamical evolution, we get the effective Hamiltonian in energy representation as follow
\begin{equation*}
H_{eff}=\sum_{k=1}^{2}\frac{\tilde{\Omega}_{k,eff}}{2}\tilde{\sigma}_z^{(k)}+\xi_k\tilde{\sigma}_x^{(k)}+\alpha \tilde{\sigma}_z^{(1)}\tilde{\sigma}_z^{(2)}+\sum_{k\neq l=1,2}\beta_{kl}\tilde{\sigma}_z^{(k)}\tilde{\sigma}_x^{(l)}+\gamma \tilde{\sigma}_x^{(1)} \tilde{\sigma}_x^{(2)},
\end{equation*}
where
\begin{equation*}
\begin{aligned}	
&\tilde{\Omega}_{k,eff}=\Omega_k \left(1+\frac{2B_k^2}{(\Omega_k-\omega_k)(\Omega_k+\omega_k)}\right),\\
&\xi_k=-\frac{A_k B_k \Omega_k}{(\Omega_k-\omega_k)(\Omega_k+\omega_k)},\\
&\alpha=\frac{2A_1 A_2 g}{\omega_1 \omega_2},\\
&\beta_{kl}=A_k B_l g\left[\frac{1}{\omega_k\omega_l}+\frac{1}{2(\Omega_l+\omega_l)(\Omega_l+\omega_k)}+\frac{1}{2(\Omega_l-\omega_l)(\Omega_l-\omega_k)}\right],\\
&\gamma=\frac{B_1 B_2 g}{2}\sum_{i\neq j=1,2}\left[\frac{1}{(\Omega_i+\omega_i)(\Omega_i+\omega_j)}+\frac{1}{(\Omega_i-\omega_i)(\Omega_i-\Omega_j)}\right].
\end{aligned}
\end{equation*}
Converting it into the location representation, we have
\begin{equation*}
H_{eff}=\sum_{i=1}^{2}\left(\frac{2}{\epsilon_i}\sigma_z^{(i)}+t_{i,eff}\sigma_x^{(i)}\right)+J_z'\sigma_z^{(1)}\sigma_z^{(2)}-\sum_{i\neq j}J_{xz,ij}' \sigma_z^{(i)} \sigma_x^{(j)},
\end{equation*}
where
\begin{equation*}
\begin{aligned}
&t_{i,eff}=t_i\left[1+\frac{2g_i^2}{(\Omega_i+\omega_i)(\Omega_i-\omega_i)}\right],\\
&J_z'=\sum_{i\neq j}\frac{2g_1 g_2 g t_i^2}{\Omega_i^2}\left[\frac{1}{(\Omega_i-\omega_i)(\Omega_i-\omega_j)}+\frac{1}{(\Omega_i+\omega_i)(\Omega_i+\omega_j)}\right]+\frac{g_1g_2g\epsilon_1\epsilon_2}{\omega_1\omega_2\Omega_1\Omega_2}\sum_{i\neq j}\frac{\Omega_i\epsilon_j}{\Omega_j \epsilon_i},\\
&J_{xz,ij}'=\frac{g_1 g_2 g \epsilon_jt_j}{\Omega_j^2}\left[\frac{1}{(\Omega_j-\omega_j)(\Omega_j-\omega_i)}-\frac{1}{(\Omega_j+\omega_j)(\Omega_j+\omega_i)}\right]+\frac{2g_1g_2g\epsilon_1\epsilon_2}{\omega_1\omega_2\Omega_1\Omega_2}\frac{\Omega_it_j}{\Omega_j\epsilon_i}.
\end{aligned}
\end{equation*}
The effective Hamiltonian here is exactly the same in form as the effective Hamiltonian describing the indirect coupling between DQDs intermediated by one Boson cavity \cite{Contreras2013Non}, both are composed of the Ising interaction term and XZ exchange interaction term, which is the origin of the similar steady transport property between the two system.
\section{Dependence of steady state of DQD on interdot tunneling}
\begin{figure}
	\includegraphics[width=12cm]{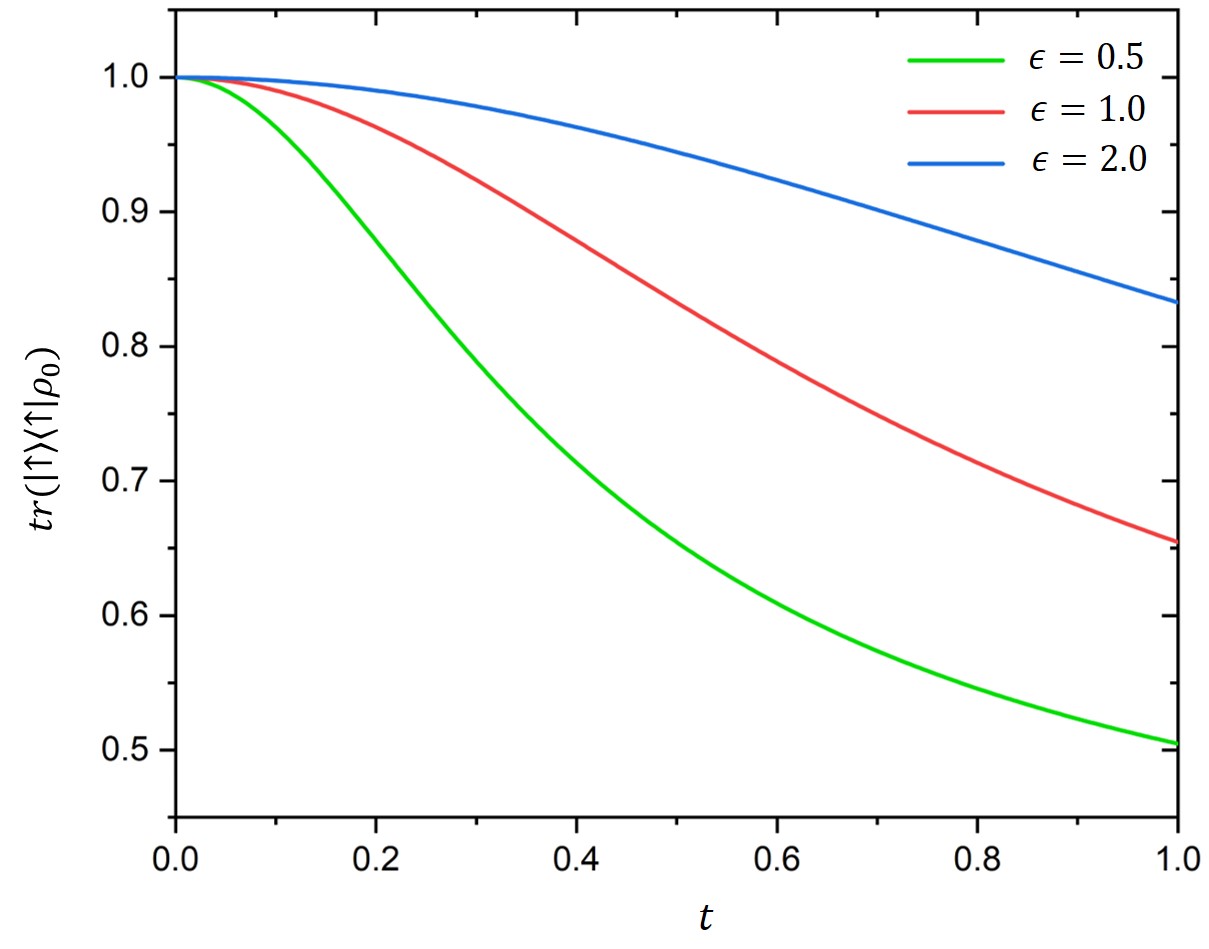}
	\caption{Simulation of $tr(\ket{\uparrow}\bra{\uparrow}\rho_0)$ versus $t$ for different $\epsilon$, here $\Gamma_1=\Gamma_2=10^{-4}$. The result clearly when $t \ll  |\epsilon|$, the steady state of DQD is almost $\ket{\uparrow}$,  $tr(\ket{\uparrow}\bra{\uparrow}\rho_0)|_{t=1}=1$, and $\frac{\partial}{\partial t}tr(\ket{\uparrow}\bra{\uparrow}\rho_0)|_{t=0}$, as we proved. \label{figure.5}}
\end{figure}

In this section, the main conclusion is that when $t\ll |\epsilon|$, the steady state of DQD isolated from resonators is almost $\ket{\uparrow}, when \epsilon>0; \ket{\downarrow}, when \epsilon<0$, which is a deduction of 
\begin{equation*}
	tr(\ket{\uparrow}\bra{\uparrow})|_{t=0}=1,
\end{equation*}
\begin{equation*}
	\frac{\partial}{\partial t}tr(\ket{\uparrow}\bra{\uparrow})|_{t=0}=0,
\end{equation*}
when $\epsilon>0$ and vice versa when $\epsilon<0$ (only need to change $\bra{\uparrow}$ with $\bra{\ket{\downarrow}}$ for the two equations above). We will demonstrate the equations above with the help of Bloch sphere. The first equation is obvious as
\begin{equation*}
	\ket{\uparrow}\bra{\uparrow}=\rho_0=\ket{L}\bra{L}, when\ t=0,
\end{equation*}
so
\begin{equation*}
	tr(\ket{\uparrow}\bra{\uparrow}\rho_0)|_{t=0}=tr(\ket{\uparrow}\bracket{\uparrow}{L}\bra{L}\rho_0)|_{t=0}+tr(\ket{\uparrow}\bracket{\uparrow}{R}\bra{R}\rho_0)|_{t=0}=1.
\end{equation*}
As for the second equation, we have
\begin{equation*}
	\frac{\partial}{\partial t}tr(\ket{\uparrow}\bra{\uparrow}\rho_0)|_{t=0}=tr\left(\frac{\partial}{\partial t}(\ket{\uparrow}\bra{\uparrow})|_{t=0} \ket{L}\bra{L}\right)+tr\left(\ket{L}\bra{L}\left(\frac{\partial}{\partial t}\rho_0\right)|_{t=0}\right),
\end{equation*}
as $\ket{\uparrow}\bra{\uparrow}$ is rotating on the surface of Bloch sphere with t, and $\bracket{\uparrow}{L}|_{t=0}=1$, so $\left(\frac{\partial}{\partial t}\bra{\uparrow}\right)|_{t=0}\ket{L}=0$, which means $tr\left(\frac{\partial}{\partial t}(\ket{\uparrow}\bra{\uparrow})|_{t=0}\ket{L}\bra{L}\right)=0$. When $t=0$, the noise and decoherence origin from $\Gamma_{L/R}$ should be limited by $t$ to zero, leading to $\left(\frac{\partial}{\partial t}|\rho_0|\right)|_{t=0}=0$, where $|\rho_0|$ means the length of the Bloch vector, on the other hand, the rotating of $\rho_0$ on Bloch sphere surface don't contribute to $tr\left(\ket{L}\bra{L}\left(\frac{\partial}{\partial t}\rho_0\right)|_{t=0}\right)$, so $tr\left(\ket{L}\bra{L}\left(\frac{\partial}{\partial t}\rho_0\right)|_{t=0}\right)=0$. And we conclude that $\frac{\partial}{\partial t}tr(\ket{\uparrow}\bra{\uparrow}\rho_0)|_{t=0}=0$. For the case where $\epsilon<0$, similar demonstration is available. Simulation of the dependence of steady state of DQD on interdot tunneling confirms our results, as shown in Fig .\ref{figure.5}.
\section{Steady current through DQD}
\begin{figure}
	\includegraphics[width=12cm]{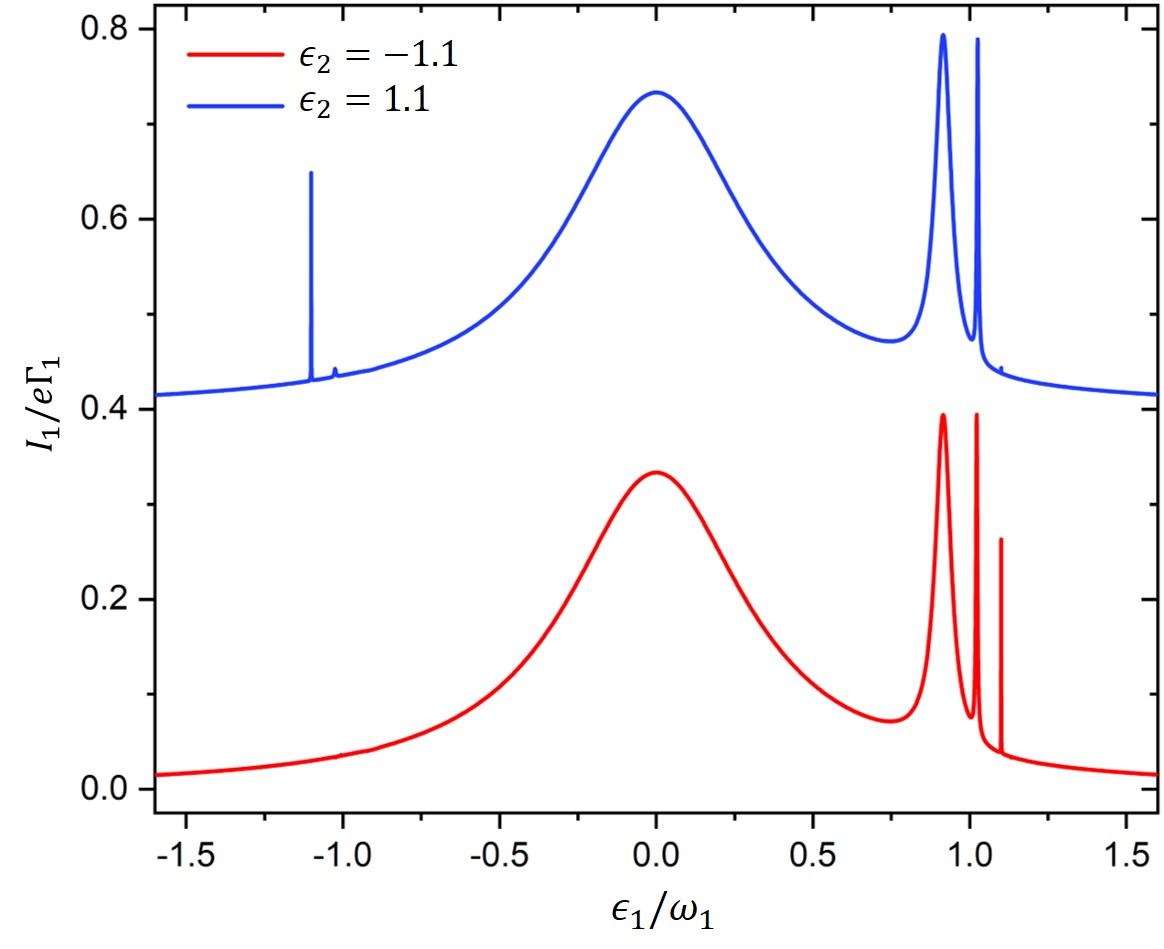}
	\caption{Simulation of $I_1$ versus $\epsilon_1$ based on the full Hamiltonian. The parameter set up is the same as that in Fig .\ref{figure.2}. The blue line is for $\epsilon_2=1.1$ with an offset of 0.4 in y-axis, while the red line is for $\epsilon_2=-1.1$. There is corresponding current peak with same $\epsilon_1$ for all concurrence peak, meaning that the steady current is qualified for detecting of concurrence peak. \label{figure.6}}
\end{figure}

By comparing steady current through DQD and concurrence of qubits, the steady current can serve as an indicator of entanglement. The simulation of steady current is shown in the Fig .\ref{figure.6}.
Two kinds of current peak are observed, which are elastic current peak at $\epsilon_1=0$, inelastic peak at resonance $\Omega_1=\omega_1,\omega_2\ or\ \Omega_2$. Noting that the peak with $\Omega_1=\Omega_2$ is distinct only in the opposite side of $\epsilon_1$ to $\epsilon_2$, which is a consequence of the XZ exchange interaction (indirect coupling) between quits.
\section{Intermediate parameter set up}
\begin{figure}
	\includegraphics[width=12cm]{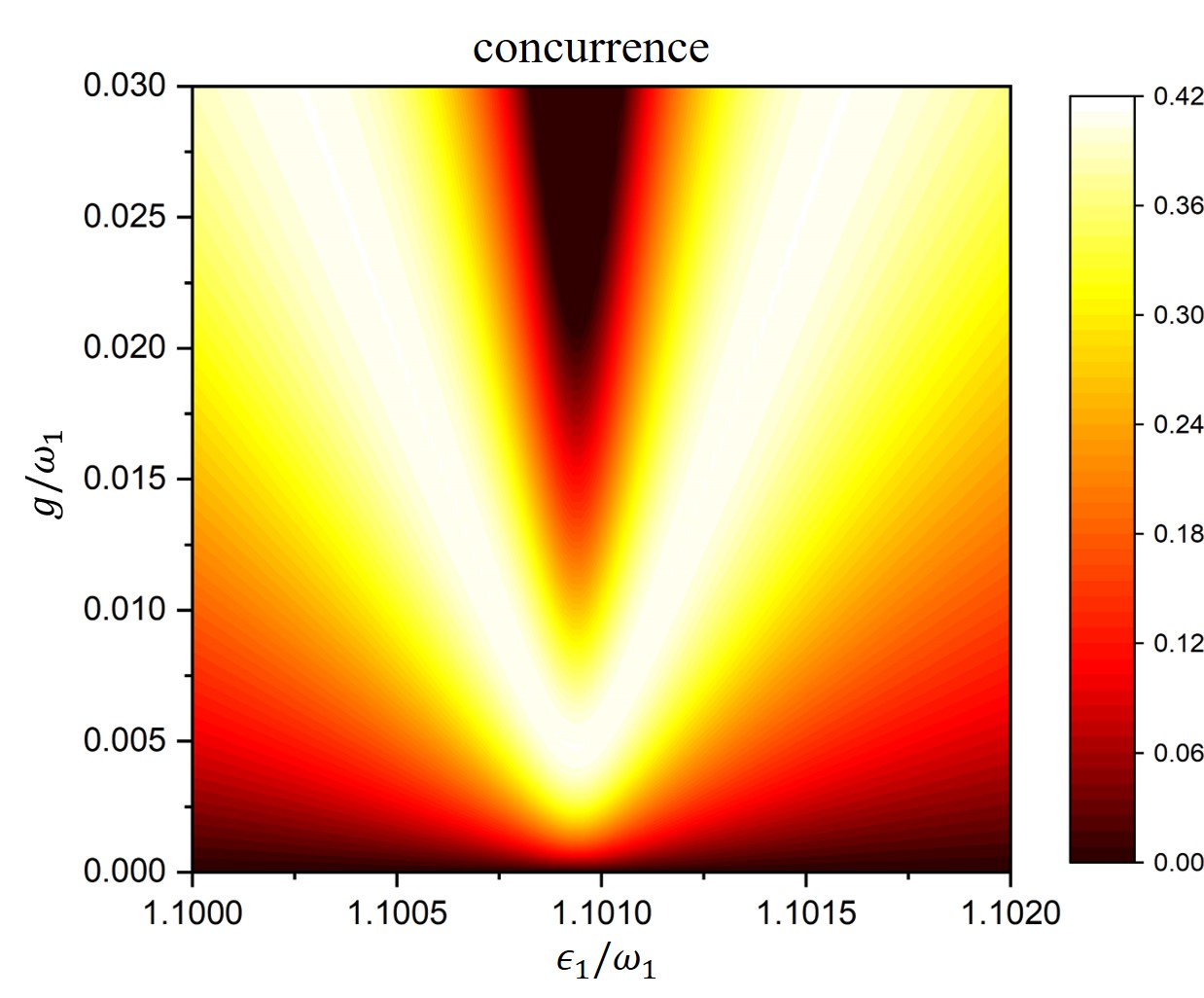}
	\caption{Steady concurrence versus $g$ and $\epsilon_1$ based on the effective Hamiltonian, with the same parameter set up as in Fig .\ref{figure.2}. A double peak structure of concurrence versus $\epsilon_1$ splitting with increasing $g$ is observed here. \label{figure.7}}
\end{figure}

To study the inermediate transition of steady concurrence from weak coupling to strong coupling, we calculated steady concurrence versus $g$ and $\epsilon_1$, shown in Fig .\ref{figure.7}. We find a double peak structure of concurrence versus $\epsilon_1$ splits with increasing $g$. The split can be explained with the deviation of eigenstates from $\ket{\uparrow\downarrow}$ taking place at a $\epsilon_1$ more distant from resonance as a result of stronger indirect coupling origins from strong $g$, leading to a descend of concurrence more distant from resonance.
\section{Eigenstates of the full Hamiltonian with RWA}
Here we give eigenstates and eigenenergy of the full Hamiltonian with RWA. The states with single stimulation are concerned. Considering the case where resonators reach resonance and qubits reach resonance respectively, in the subspace $span(\ket{\uparrow\downarrow 00},\ket{\downarrow\uparrow 00},\ket{\downarrow\downarrow 10},\ket{\downarrow \downarrow 01})$, the reduced Hamiltonian reads
\begin{equation*}
H=
\begin{pmatrix}
\Omega&0&B&0\\
0&\Omega&0&B\\
B&0&\omega&g\\
0&B&g&\omega
\end{pmatrix}
,
\end{equation*}
where $B_1=B_2=B$. The eigenenergy is
\begin{equation*}
	E_{1,2}=\frac{1}{2}\left(\omega+\Omega-g\pm \sqrt{(\Omega-\omega+g)^2+4B^2}\right),
\end{equation*}
\begin{equation*}
E_{3,4}=\frac{1}{2}\left(\omega+\Omega+g\pm \sqrt{(\Omega-\omega-g)^2+4B^2}\right),
\end{equation*}
corresponding to eigenstates $(B,B\eta_i,E_i-\Omega,(E_i-\Omega)\eta_i)'$, where
\begin{equation*}
	\eta_1=\eta_2=-1,\eta_3=\eta_4=1.
\end{equation*}

\begin{acknowledgments}
This work is supported by National Key Research and Development Program of China (2018YFA0306102, 2018YFA0307400); National Natural Science Foundation of China (91836102, 61704164, 12074058).
\end{acknowledgments}

\bibliographystyle{apsrev4-2}
\bibliography{reference}

\end{document}